\documentclass[usenatbib]{basi}
\pdfoutput=1
\usepackage[T1]{fontenc}
\usepackage[british]{babel}
\usepackage[varg]{txfonts}
\usepackage{rotating}
\usepackage{dcolumn}
\begin{document}
\title[Origins of Radio Astronomy at TIFR]{Origins of Radio Astronomy at the Tata Institute of Fundamental Research and the Role of J.L. Pawsey}
\author[W.M. Goss]%
       {W.M. Goss$^1$\thanks{email: \texttt{mgoss@nrao.edu}}\\
       $^1$NRAO, PO Box 0, Socorro, New Mexico 87801 USA}

\pubyear{2014}
\volume{00}
\pagerange{\pageref{firstpage}--\pageref{lastpage}}
%\status{submitted}

\date{Received --- ; accepted ---}

\maketitle
%------------------------------------------------------------------------------%
% abstract and keywords                                                        %
%------------------------------------------------------------------------------%
\label{firstpage}

\begin{abstract}
\end{abstract}
I will discuss the interactions of a number of individuals that played major roles in the formation of radio astronomy in India in the period 1952-1962, particularly Dr. Joseph L. Pawsey.  The story began in 1953-1954: Pawsey brought Govind Swarup to Australia as a Colombo Fellow in 1953, where he worked with Christiansen, Mills, Wild and Bolton.  Later, Swarup went to Stanford where he completed a PhD with Ron Bracewell working on the new Solar Microwave Spectroheliograph. In the era 1960-1963, with the encouragement of Pawsey, several colleagues in Australia and Bracewell, discussions began among a number of Indian colleagues to form a radio astronomy group in India.  The main players were G. Swarup, T.K. Menon, M.R. Kundu and T. Krishnan.  Homi J. Bhabha, the Director of TIFR, made the decisive offer to this group to start a radio astronomy project in early 1962.  Swarup joined TIFR in early April 1963.  Many factors contributed to the successful formation of the new group: international networking among scientists of several generations, rapid decisions by Bhabha and the readiness to take chances in choosing promising, young, energetic scientists.  In December 2013, we have celebrated 50 years of ground breaking research by the TIFR radio astronomers as well as the outstanding decade of research with the GMRT- the Giant Metrewave Radio Telescope.  Govind Swarup has provided the inspiration and leadership for this remarkable achievement.


%------------------------------------------------------------------------------%
% main text of the paper, using \section, \subsection, \subsubsection          %
%------------------------------------------------------------------------------%
\section{Introduction}\label{s:intro}

In their contributions to this volume, Govind Swarup and Ron Ekers have summarised the history of radio astronomy at TIFR.  Ekers has also highlighted Govind Swarup's major role in the success of TIFR radio astronomy.  In this contribution I will provide additional details of the Australian-Indian connection in the mid 20th century\footnote{The connections of Pawsey with the formation of the TIFR radio astronomy will also be summarised in a future biographical study of J. L. Paswey by W. M. Goss, Claire Hooker and R. D. Ekers. References are based on the archive of the National Centre for Radio Astrophysics, TIFR, Pune, except where otherwise noted in the footnotes.}.  Many prominent astronomers and physicists have played direct and indirect roles in the establishment of radio astronomy at the Tata Institute of Fundamental Research in 1963.  As an example, two of the major personalities were Homi Bhabha and Joe Pawsey, both students at the Cavendish Laboratory in Cambridge in the 1930s.  Numerous scientists contributed to the success of the formation of TIFR radio astronomy over a several decade period:  J.L. Pawsey, W.N. Christiansen, R.N. Bracewell, K.S. Krishnan, H. J. Bhabha, Govind Swarup, T.K. Menon, Mukul R.Kundu, T. Krishnan, Mark Oliphant and Bart Bok.  In this contribution, I will provide brief background information about Pawsey and Bhabha, followed by Govind Swarup's decisive research experience in Australia in the 1950s followed by two appointments in the US (Harvard and then Stanford).  Some details of the negotiations that lead to the formation of the TIFR radio astronomy group during 1961-1963 will also be provided.  

\section{J. L. Pawsey (1908-1962) and H. J. Bhabha (1909-1966)}
\subsection{Pawsey}
Pawsey was born in Ararat Victoria (Australia) on 14 May 1908.  He received his BSc degree (Physics) in 1929 from the University of Melbourne, with a MSc in 1931.  No university in Australia awarded a PhD degree in Physics until after World War II.  Pawsey went to the University of Cambridge in the UK to work with J.A. Ratcliffe (1902-1987) at the Cavendish Laboratory from 1931-1934, with a PhD thesis on the properties of the E layer of the ionosphere.   Pawsey worked on early television problems at E.M. I. Electronics Ltd in the UK from 1934 to 1939; he then left the UK for Australia with his young family following the outbreak of WWII.  He began work at the Radiophysics Laboratories (RP) of CSIR in February 1940, working on radar projects until 1945.  During the war, he and Ruby Payne-Scott carried out an early radio astronomy observation at 11 cm in March 1944 (Goss, 2013). 

 The major achievement of Pawsey's career was the establishment and leadership of the Australian radio astronomy research group in October 1945, beginning with ground breaking observations of the sun using surplus war time radar receivers with colleagues Ruby Payne-Scott and Lindsay McCready (Pawsey, Payne-Scott and McCready 1946).  During the war and immediately afterwards, Pawsey recruited and mentored many of the pioneers of early radio astronomy including Mills, Bracewell, Wild, Kerr, Christiansen, Payne-Scott and Bolton. 

\subsection{Bhabha}
Homi J. Bhaba was one of the major   scientific personalities of India in the 20th century; he is known among scientists and the Indian public as ``the Father of India's Atomic Energy Programme'' (Sreekantan, 1998).  He and Pawsey were at Cambridge at the Cavendish Laboratory at the same time; Bhabha's PhD was awarded in 1935 (advisor R.H. Fowler) titled ``On Cosmic Radiation and the Creation and Annihilation of Positrons and Electrons''.  During the 1930s, he became ``famous for his 'cascade theory of the electron'. The research that he did during this period had a direct bearing on the resolution of several important issues of cosmic ray phenomena and the interactions of particles especially electrons, protons and photons at high energies, in the context of the developments in the field of quantum mechanics and relativity'' (Sreekantan, 1998).  He departed Cambridge in 1939 for the Indian Institute of Science (Physics Department, headed by C.V. Raman, 1888-1970).  In 1945, he left the Indian Institute of Science to found the Tata Institute of Fundamental Research in Bombay.  This institute ``...under his leadership [became] a major centre of cosmic ray research covering all aspects of the radiation and continues to be active in the field'' (Sreekantan, 1998).

 Numerous biographies of Bhabha have been written, including the magnificently illustrated {\it A Masterful Spirit}, Homi J. Bhabha 1909-1966 by Indira Chowdhury and Ananya Dasgupta, published in 2010.  In addition Govind Swarup has written a short biographical sketch in 1991 in {\it Current Science} ``Homi Bhabha - personal reminiscences\footnote{A well-known scientific biography was published by G. Venkataraman in 1994: {\it Bhabha and his Magnificent Obsessions}.}''.  A fascinating interview with B.V. Sreekantan (TIFR Director 1975-1987) carried out by Rica Malhotra ``Homi Jehangir Bhabha: A Visionary'' was published in Resonance in May 2010.  A photograph of Bhabha is shown in Fig.3 in the paper by Swarup in this volume.  Tragically, both Pawsey and Bhabha died prematurely in the prime of their careers. Homi Bhabha died in an Air India crash in the Alps (near Mount Blanc) on the way to Geneva and Vienna (meeting of the Advisory Committee of the International Atomic Energy Commission) on 24 January 1966. ``[I]f he had lived for another 20 years, [India] would have progressed very much.  There is no question about it absolutely because he had the vision and he had the capacity and the Government's support'' (Sreekantan, 2010)\footnote{Sreekantan (2010) has pointed out a major trait of Bhabha's that had lasting consequences:'' One [...] very important thing he did in the early years of development was making abundant use of the international contacts he had. We had a stream of foreign visitors and experts in various fields giving lectures, participating in discussions and criticising our programmes when they were not right. All this helped in evaluating our activities and laying a pathway for the future.'' The radio astronomers benefited from this phenomenon in the years after 1960.}. Pawsey was to die at age 54 at the end of 1962, (See section 4.6, footnote 12).

\section{URSI Sydney 1952}

\begin{figure}[h!]
\centerline{\includegraphics[width=9cm]{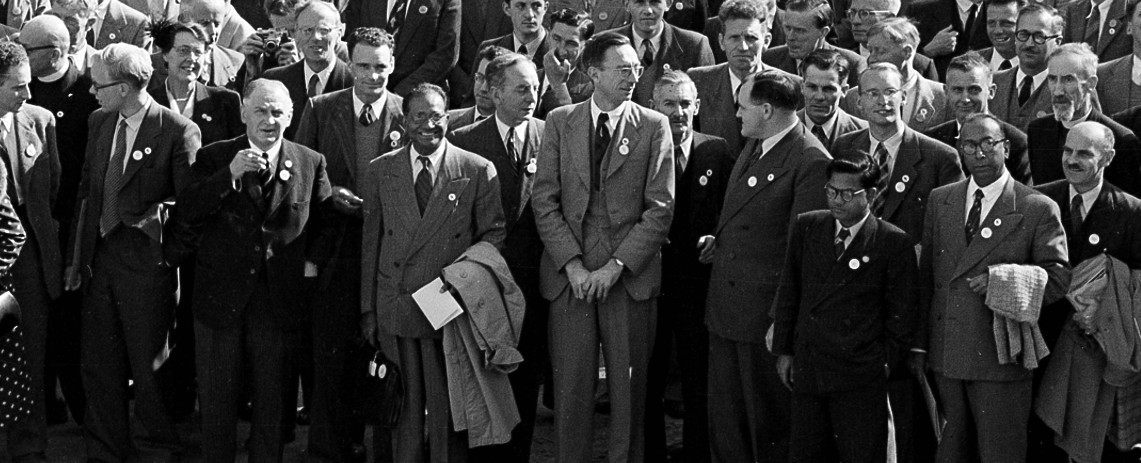}}
\caption{ Delegates at the URSI General Assembly at Sydney, 1952 included Sir. K. S. Krishnan (middle), Prof. S. K. Mitra, and Dr. A. P. Mitra.\label{f:one}}
\end{figure}

An important event for Australian science occurred in 1952 when the International Union of Radio Science (URSI) held its 10th General Assembly at the University of Sydney (Goss and McGee 2009).This was one of the first international scientific assemblies to be held outside Europe or North America.  The congress was a vehicle to showcase the remarkable achievements of radio science (in particular radio astronomy) in Australia, which had occurred in the seven years since the end of WWII.  The three Indian delegates were Sir K. S. Krishnan (1898- 1961), Director of the National Physical Laboratory in New Delhi, Prof. S.K. Mitra (1890-1963), Director of the Institute of Radio-physics and Electronics of the University of Calcutta and A. P. Mitra (1927-2007), then a young physicist from S.K. Mitra's institute. A.P. Mitra was a long term visitor at CSIRO Division of Radiophysics in Sydney, working on ionospheric projects with Alex Shain.  These three plus Pawsey and Bowen (the Chief of the CSIRO Division of Radiophysics, RP) and E. V. Appleton are shown in the group photograph during the URSI Congress at the University of Sydney in Fig. 1.

\section{ Govind Swarup (23 March 1929-)}
Swarup has discussed the beginning of his career in his contribution to this volume, summarising a remarkable saga.  (see  Swarup, 2006, ``From Potts Hill -Australia- to Pune-India: The Journey of a Radio Astronomer'').  After obtaining his M.Sc. degree from Allahabad University in 1950, he joined the National Physical Laboratory under the direction of K.S. Krishnan, working on problems in paramagnetic resonance.  Swarup was taught a course on electricity and magnetism by Krishnan at the Allahabad University during his B.Sc degree in 1946-1947.  Krishnan ``was struck by the dramatic and remarkable discoveries being made in the field of radio astronomy by staff [from Radiophysics in Sydney].  Under the inspired leadership of J. L. Pawsey several ingenious radio telescopes had been developed by Australian scientists to investigate radio emission from the Sun and distant cosmic sources...On his return to India, Krishnan described these developments in a colloquium at the NPL, and these caught my [Swarup's] imagination''.  Swarup went to the library and read the Australian scientific papers:  ``I was told that these were almost half of the papers on radio astronomy that had been published worldwide up to that time.  I, too, was fascinated by this new field.  Krishnan was also interested in initiating radio astronomical research at the NPL, and he put my name forward for a two year Fellowship under the Colombo Plan to work at [Radiophysics] in Sydney'' (Swarup, 2006). Swarup was only 22 years old when he ``first met Bhabha during his visit to the National Physical Laboratory (NPL) in New Delhi in 1951.  He was actively participating in the launching of an array of balloons for cosmic ray-research.  His drive and attention to details were awesome and infectious'' (Swarup, 1991).   

\subsection{Swarup in Australia 1953-1955}

\begin{figure}[h!]
\centerline{\includegraphics[width=5cm]{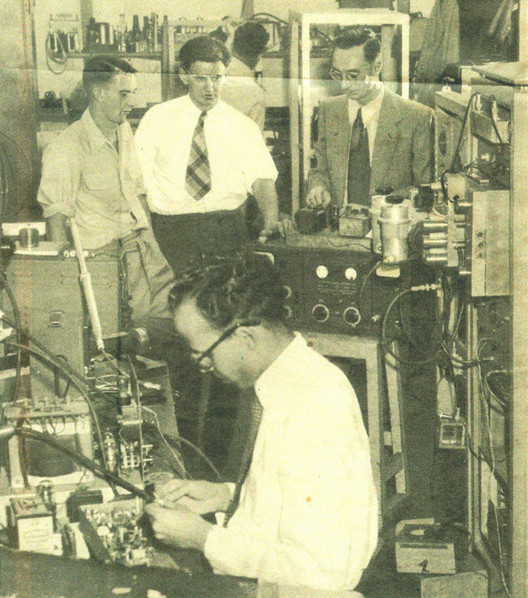}}
\caption{ Govind Swarup with J. L. Pawsey, J. G. Bolton and G. Stanely (right to left) at the CSIRO Radiophysics lab. \label{f:two}}
\end{figure}

In March 1953  the two radio astronomy recipients of Colombo Plan
fellowships, Govind Swarup and R. Parthasarathy (from the Kodaikanal
Observatory in Tamil Nadu) arrived at CSIRO to work with Pawsey.  At
Radiophysics, Swarup had a remarkable career during the two year four
months period working in turn for three month periods with
W.N. (``Chris'') Christiansen (1913-2007), J. Paul Wild (1923-2008),
Bernard Y. (``Bernie'') Mills (1920-2011) and John G. Bolton
(1922-1993).   A picture of Swarup in the laboratory of Bolton and
Stanley is shown in Fig. 2, a photograph from a popular Sydney
magazine (People) from 10 February 1954: ``Stargazers- Australian
Radio Astronomers- J.L. Pawsey and his Group Lead the World in this
New Science''.  Unfortunately this photo has often been shown with Swarup's image cropped!  In the magazine, Swarup was not even identified, only the three more senior colleagues Pawsey, Bolton and Gordon Stanley (1921-2001).   An important event occurred during the first part of Swarup's visit to Sydney: Bhabha was visiting Australia and was given a tour of the Christiansen Potts Hill grating array by Pawsey.  ``In 1955, [Bhabha] visited Potts Hill station of CSIRO near Sydney, where many dramatic and remarkable discoveries had been made by [RP staff] under the leadership of J.L. Pawsey'' (Swarup, 1991). The major lasting influence experienced by Swarup was working with Chris Christiansen: ``I learned the powerful technique of radio interferometry from Chris in 1953 and have not looked back'' (Swarup, 2008). In this volume, Ron Ekers has described the role played by Swarup in calculating the Fourier transforms by hand for the observations published by Christiansen and Joe Warburton (1955).  Govind Swarup was not a co-author (he has told the author that Warburton had insisted on this condition) and was given an acknowledgement at the end of the paper.  In Fig. 10 of Christiansen and Warburton (1955), the resultant image of the sun at 20 cm with a 4.3 arc min resolution is shown, after the SVC (slowly varying component- the enhanced emission) had been subtracted.  In his contribution to this volume, Swarup (his Fig. 1) has presented an image of the Potts Hill grating array.  The major result from the quiet sun data was the determination of limb brightening in the equatorial zones of the sun with none in the polar regions.  Earlier Stanier (1950) had found no evidence for limb brightening at 60 cm using observations from Cambridge.  The Potts Hill image of the corona was in good agreement with the optical images of the corona taken during eclipse at this epoch of solar minimum.  

In 1954, Christiansen went to Meudon in France for a year to work with the French solar astronomers.  Back in Sydney, Swarup and Parthasarathy (under the direction of Pawsey) converted the Potts Hill EW array of 32 dishes to 500 MHz to investigate the possible limb brightening at 60 cm. (``It took myself and Parthasanathy only three months to convert Chris's interferometer and receiver to 60 cm [from 20 cm] including the design and adjustment time.\footnote{Letter from Swarup to Bracewell on sabbatical University of Sydney, 13 December 1961.}'') Their data showed clear limb brightening at the longer wavelength of 60 cm, in contrast to Stanier's results. A summary paper for the Observatory and two detailed publications for the Australian Journal of Physics were prepared from these observations made during 1954-1955 (Swarup and Parthasarathy, 1955a, 1955b and 1958).  This experience was of great value to these young scientists, working independently on hardware construction, observations, data reduction and the interpretation of the results.  

Before Swarup left Sydney, he asked Pawsey (Assistant Chief of RP) and later on Bowen (Chief of the Division of Radiophysics) if 32 of the 1.7 m dishes (without the 16 dishes of the north-south arm) could be donated to the National Physical Laboratory (NPL) in India after the observations were to cease in late 1955. K.S. Krishnan had agreed to this donation with the condition that the Australians were to pay the modest shipping costs (\pounds 700).  The NPL was to start a modest radio astronomy programme.  Swarup had planned a grating interferometer with a baseline of 640m for solar observations using the 32 dishes at 500 MHz and 165 MHz.  The proposal ``Proposed Investigation: Study of Solar Radiation Using a 2100 foot Long 32-Element interferometer to Operate Simultaneously at 60 cm and 1.8 m Wavelength''  was sent to K.S. Krishnan on 23 January 1955 from Sydney.  The total cost was to be 10,000 rupees, about \$2100 US in 1955.

\subsection{ Back at NPL in India- 1955}

Swarup returned to NPL in July 1955; he began to build a 500 MHz receiver for the new grating array.  However, Krishnan could not get permission to obtain the foreign exchange for the shipping costs.  (Eventually CSIRO did pay for the shipping costs.) Swarup was frustrated with the delays and decided in late 1955 to go to the US for a year or two.  He ended up remaining at the Harvard Fort Davis Texas radio astronomy station for one year and 5.5 years at Stanford.

During the mid-1950s, a number of other young scientists were also associated with NPL and the nascent radio astronomy program.  All were to play significant roles in the early years of radio astronomy in India: N.V.G. Sarma  (1931-2008; Leiden,  Benelux Cross), M.R. Kundu (1930-2010, PhD, Paris), T. Krishnan (1933- ; Cambridge,  the Cavendish group of Ryle) and M.N. Joshi (1933-1988, PhD Paris). Swarup (2006) has characterised this era: ``Thus, it may be said that the NPL acted as a foster mother for the subsequent development of radio astronomy in India [since these young scientists had been] trained across the world.''

\subsection{West Texas: 1956-1957}
\begin{figure}[h!]
\centerline{\includegraphics[width=4cm]{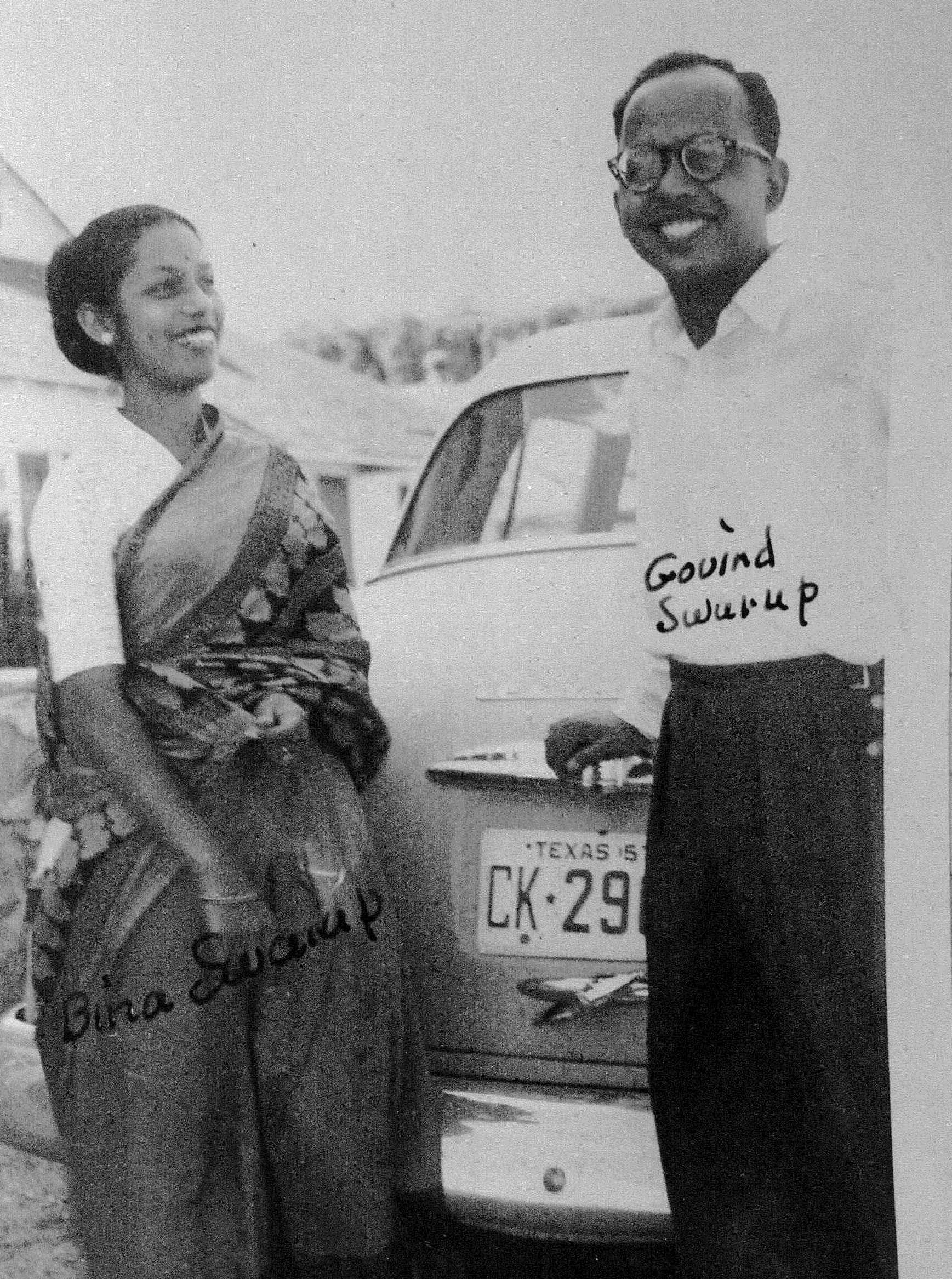}}
\caption{ Govind and Bina Swarup in West Texas.\label{f:three}}
\end{figure}

In August 1956 the newlywed family of Govind and Bina Swarup set out for the wilds of west Texas, at the Harvard Radio Astronomy Station.  Govind had been invited by Donald Menzel of the Harvard Observatory to work with Alan Maxwell on dynamic spectra of solar bursts in the range 100-600 MHz using the newly commissioned 8.5 m telescope.  A major discovery was made by Swarup in December 1958, the type U bursts (Maxwell and Swarup, 1958).  An image of Govind and Bina Swarup, in front of their car with a Texas registration, is shown in Fig. 3. 

In 1957, Swarup made an important decision, planning for a PhD degree in the US.  He was admitted into Harvard, Caltech and Stanford, all with newly formed radio astronomy groups.  On 8 February 1957, Pawsey had been asked by Swarup to provide letters of reference for admission to the graduate programs.  He replied: ``If you are returning to India, I should recommend to you to place great emphasis in electronics.  It is a key to open many doors.  Stanford is famous for radio engineering, Caltech for its physics and, of course, its astronomy research, and Harvard for its training in astronomy.'' Swarup chose to work with Bracewell at Stanford.

\subsection{ Stanford: 1957-1963}
In September 1957, Govind and Bina Swarup moved to Palo Alto, where he spent 3.5 years as a graduate student and two years on the faculty as an Assistant Professor.  An intense period of collaboration with Ronald N. Bracewell (1921-2007) began. During this period, Swarup worked on images of the sun at 11 cm from the newly completed Stanford Microwave Spectroheliograph at the Stanford site ``Heliopolis'' (first operational in April 1960, in full time operation from June 1962 to August 1973, the Heliopolis Observatory closed in 1979, see Bracewell, 2005). A photo of the Stanford Microwave Spectroheliograph is shown in Fig. 2 of Swarup's contribution in this volume. As a part of his PhD (end 1960, with Bracewell as supervisor) research, Swarup found a north-south asymmetry in the 9.2 cm emission of the quiet sun. Pawsey was impressed (18 April 1961): ``I am most interested in your ... report.  I am hoping it is the essentials of a PhD thesis.  I thought it was a most interesting report and I congratulate you on the outcome of your Stanford work.  Your report now will be my authoritative source on the features of 10 cm observations.''  The definitive publication on the new instrument appeared in 1961: ``The Stanford Microwave Spectroheliograph Antenna, a Microsteradian Pencil Beam Interferometer'', Bracewell and Swarup, 1961.  In this paper from 1961, Fig. 7 is an image of the sun on 30 May 1960 with a resolution of 3.1 arc min at 9.2 cm.

A major achievement was the publication of the phase measurement scheme of a long transmission line due to temperature changes.  Swarup had worked on this problem at Potts Hill in the 1950s; he had proposed a scheme at that time which Pawsey pointed out could not work (Swarup, 2008): ``Six years later [at Stanford], I took care of Pawsey's objections by conceiving a round trip phase measurement scheme and modulating the signals at the output of the Stanford array parabolic dishes.  Hence, I conceived the idea of transmitting a signal at 9.2 cm from a central point of the transmission line to all antennas, modulating and reflecting the voltage signal from the output of each antenna and measuring the round trip phase of the modulated signal''.   ...(Swarup, 2008). The method was described by Swarup and Yang (1961), a scheme now widely used for phase adjustments to many   telescopes of the last 50 years (and many other applications).  Pawsey was impressed, writing on 26 October 1960: ``I had already heard of your phase measurement technique and think you have made a real break-through in this technique.  Congratulations! Chris [Christiansen] regards the idea as the key to really large Mills Crosses.  Without a good checking technique they could not operate.  We are planning a new equipment (compound interferometer with one arm of the Chris-Cross [at Fleurs] and a 60-foot dish) using this technique.''

\subsection{Planning for a Radio Astronomy Group in India: 1960-1961}

\begin{figure}[h!]
\centerline{\includegraphics[width=7cm]{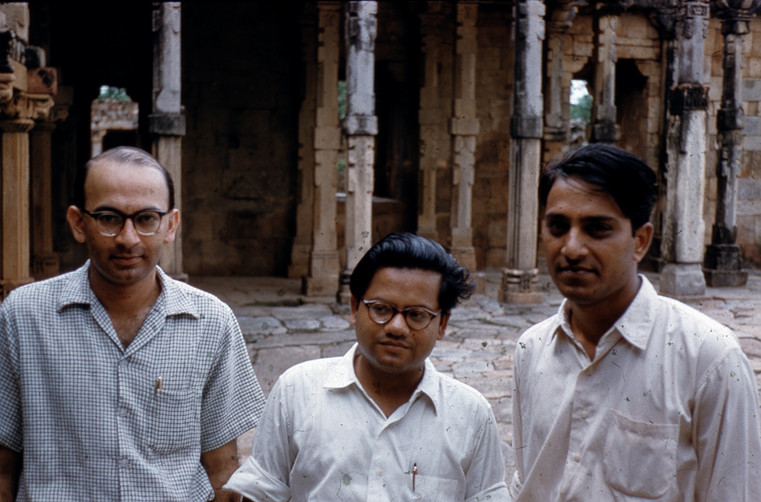}}
\caption{ J. L. Pawsey's photograph of T. Krishnan, A. P. Mitra and N. V. G. Sarma (left to right) at NPL, New Delhi \label{f:four}}
\end{figure}

A few months before Swarup's PhD degree was completed, he wrote Pawsey on 29 September 1960.  He explained that he would like to remain at Stanford for a year or two and then return to India.  Along with M. (Mukul) R. Kundu and  (Kochu) Menon, discussions had started about a new Indian radio astronomy group.  ``We might be able to combine our efforts to start a radio-astronomy group somewhere in India.'' Swarup had just met Dr. B.D. Laroia, Development Officer of the University Grants Commission of India, who had suggested that a major university in India might start a radio astronomy group.  Ron Bracewell suggested that the four Indian radio astronomers working in the US might meet at the next American Astronomical Society meeting to discuss this joint effort.  Swarup asked Pawsey for advice: ``It would be of great value to have your advice as to the nature, objective and location of this group in India.  It would also be of value if you stop at Delhi for a day during any future trip around the world.''  In fact, Pawsey had stopped in New Delhi in August or September 1958 after the Paris Symposium and then the International Astronomical Union General Assembly in Moscow.  Fig. 4 shows an image found by Goss in the Pawsey family slide collection in Sydney.  Pawsey visited NPL and met three of the young scientists, two who had worked with him in Sydney, T. Krishnan (see below) and A.P. Mitra (later NPL Director from 1982-1986 and  CSIR Director General from 1986 to 1992). N.V.G.Sarma (also in the photograph) was a major contributor to the RF electronics of the Ooty Radio Telescope in the early years of the TIFR radio astronomy group, starting in 1964. He moved to the Raman Research Institute in Bangalore starting in 1976. Here he made one more major contribution to the TIFR group; the L-band feeds for the nascent GMRT in 1998 were constructed in the new lab he had established.

The correspondence between Swarup and colleagues in Sydney (Pawsey, Christiansen and Frank Kerr) in the period September 1960 to March 1962 (the year of Pawsey's death ) is a fascinating record of the successful foundation of the TIFR radio astronomy group starting in April 1963 with Swarup's arrival at TIFR, Bombay, four months after Pawsey's death on 30 November 1962.  On 22 September 1960, Christiansen (newly a Professor of Electrical Engineering at the University of Sydney) was about to travel to Leiden for year's visit to work on the Benelux Cross.  He wrote Swarup: ``It seems that you two [Swarup and T. Krishnan- see below] and Menon and Kundu should get together for a united attack on the monolith of Indian bureaucracy- separately I can't see you getting anywhere in radioastronomy very fast.  ...I think you should do something fairly soon.  I know you all, and feel that [this group of four] would make a very fine team.''

A month later Pawsey (26 October 1960) continued the theme: ``I think your idea of getting a group of Indians like yourself together to start a new radio astronomy laboratory is very sound.  To my mind it is much the most promising possibility for making India play an effective role in radio astronomy... With regard to your nucleus, you mention Kundu and [T.K.] Menon plus yourself.  An important potential addition is T. Krishnan who is now with us [at CSIRO in Sydney] but plans to leave next June\footnote{In fact T. Krishnan moved to Stanford to work with Bracewell in January 1963, close to the time that the Swarup family left the US for India, 27 February 1963, via Europe and then via ship from Italy arriving in Bombay on 30 March 1963.}. ...Krishnan has been very interested in just this possibility...He has friends in high places and his influence could be a great help.'' Then Pawsey sounded a prescient warning: ``I should like to add a warning here.  It will probably happen that different ones want different things.  You must all try to sink your personal preferences in favour of the whole project and judge objectively.  Remember that strength lies in unity.  I have wondered at the possibility of a meeting of the group in Berkeley in August next (IAU)... I would favour an organization associated with, but not an integral part of, a university. ...The main thing is enthusiastic support... . But keep off fashionable stuff as far as possible.  Be original.  Try, if possible, to develop ideas which one or more of you have originated.  The small groups bring in the radical new thoughts.  They are not tied up with inherited large programmes.  The other point for me to emphasize is the importance of good experimental technique.  This is the usual weak point in Indian groups, so do your best to beat this from the start.''

Some months later (18 April 1961), Pawsey continued his warning to Swarup: ``If the scheme comes off, you are a key person.  In general, Indian science tends to be stronger on the theoretical side than the practical side and to be not sufficiently critical and single minded.  Your practical ability and direct approach will be most important... A real difficulty is the question of leadership of the group.  It could well lead to jealousies and failure.  The position is that there are several of you, each with something different to contribute, and no one alone is likely to be able to make things go by himself... .  [The] problem... .  can be met if the individuals of the group are each willing to subordinate their individual interests to some extent... My reaction is that you take a good look at what is going on in the US and then do something different.  Don't try to compete with big money, but try to be original.  It is easy to say but hard to realize.''

Pawsey continued his advice on 29 June 1961, clearly worried about management issues: ``...you, I think stand out in practical experience and ability.  Menon has the astronomical knowledge.  Who then will be the leader? This is the sort of situation which can be made to work if the members have a real urge to make the project a success and are willing to subjugate their individualities to a reasonable extent.  Not all men can do this and you do not want to pick up members who won't fit in.  On the other hand, you must get a few key people.  My own feeling is that there are only two essential key men: (1) a good practical physicist combing radio skill and common sense (you are my choice here); and (2) an organizer with drive and good external contacts (Krishnan stands out   here).''  Again he urged originality: `` ...try to avoid one of the 'fashionable rackets' of the US.  Don't, for example, just buy a 60 foot dish because someone gives it you cheap.  America is stiff with 60 foot and 85 foot dishes bought by organizations who had no special ideas of what to do with one.''

On 25 July 1961, Swarup responded to Pawsey mentioning additional Indian colleagues that might well play roles in the new radio astronomy institute, likely to be a part of an existing institution in India rather than a new institute.  Besides the initial list of four (Swarup, T.K. Menon, Kundu and Krishnan), others mentioned were V. Radhakrishnan\footnote{Radhakrishnan was still interested in working overseas for a few years. He wrote Swarup on 7 January 1962 while on leave from Caltech at Bell Labs: ``Do keep me informed of the progress as it seems inevitable that sooner or later I shall find myself helping you chaps in trying to do radio astronomy in India.'' This did occur, beginning at the Raman Research Institute after 1971; numerous radio astronomy initiatives, including major collaborations with TIFR radio astronomers, were completed in the next decades under the direction of Radhakrishnan.} (1929-2011), Sarma and Joshi.  Swarup explained that he had little interest in management of the group: ``It would be desirable for me to devote all my limited energies to the construction and development of the apparatus.  I personally would not care who would be the leader.  But the overall scientific program should be mutually decided so as to make everybody in the group feel his importance.'' He foresaw a concentration on a single scientific programme.  ``I think we do not need to worry about a leader at present but should decide on a line of attack required to establish the group.  The final arrangement would depend on those who are most motived and interested in working out dirty details, on the scope of the project and most important, on the sponsoring organization in India.'' 

A lucid text, explaining the conditions surrounding the possible formation of the new group was sent by Swarup to T.K.Menon in an undated letter (likely mid 1961): ``The question of establishing a radio astronomy group in India has been talked about for a long time and I am sure it would be discussed at the [IAU in Berkeley in August 1961].  I am certain that some kind of radio astronomy project would be definitely started in India within the next 2 or 3 years, and I do not mean a Yagi antenna and a 100 mc receiver! I say this merely because there are so many of us who are interested in this subject.  It is upon us to {\it at least attempt} [his emphasis] to plan this group in such a location and surroundings that it could grow with vigour in future if necessary and at the same time contribute to Indian science and education... .  I think that we have amongst us a very nice nucleus for an active research group in a new vigorous field of science.  The main thing in my opinion is the availability of so many Indians trained in a narrow field of science who can make a homogenous group, a situation of definite advantage in a country like India.''

During the IAU in August 1961, Krishnan, Menon, Swarup and Kundu met during the General Assembly, discussing the plan to return to India with the formation of a radio astronomy group. Within the next month, a 3.5 page proposal was completed on 23 September 1961 ``Proposal for the Formation of a Radio Astronomy Group in India'' by T. Krishnan, M.R. Kundu, T.K. Menon and G. Swarup.  ``Almost every major nation in the world has recognized the importance of these contributions for the future development of scientific and technological research by embarking on large research programs in radio astronomy.'' The most efficient organisation was to place the group ``as part of an already existing institution such as a national laboratory, university or higher technical institute.  This will make it possible for the members of the group to take an active interest in related branches of science, and in the teaching and training of students.'' An initial project was to be a modest solar telescope using the 32 1.7 m dishes that the National Physical Laboratory had received a decade earlier from CSIRO. The proposal was sent to the Government of India through the Atomic Energy Commission (also the Tata Institute for Fundamental Research, H. J. Bhabha), the CSIR (Council of Scientific and Industrial Research, M.S. Thacker), Ministry of Natural Resources and Scientific Research, Director General of Observatories, the University Grants Commission (D.S. Kothari) and the Physical Research Laboratory (Ahmedabad, Vikram Sarabhai).  The proposal also contained a section describing larger, future projects (such as a low frequency Mills Cross at 25 MHz with 10s of arc min resolution), which would be considered after the group was settled in at TIFR.

Five well distinguished astronomers outside India were sent the proposal for commentary: Jan Oort (Leiden, the Netherlands), Bark Bok (Mt. Stromlo, Australia), J.L. Pawsey (CSIRO, Australia), J.F. Denisse (Observatoire de Paris, France) and Harlow Shapley (Harvard,USA) .  The responses to Bhabha from Oort, Bok and Pawsey are available in the TIFR archives.  

Pawsey's response from 6 October 1961 is included in the book by Chowdhury and Dasgupta.  Pawsey was well placed to provide an assessment of the proposal since two of the scientists (Swarup and Krishnan) had worked with him in Sydney.  ``I have a very high opinion of the scientific talent in this group... . a group chosen from among them should have an excellent chance of building up a first class scientific institution... . I regard this spontaneous movement among the young Indians who have initiated this proposal as a most encouraging sign and strongly urge you, in the interests of science in India, to try to assist them in their efforts to work out something worthwhile.'' Bok was equally enthusiastic (23 October 1961): ``Here is a case of four young, but renowned, Indian Radio Astronomers, all thoroughly trained and with good research records... . hardly equalled by any group that one might assemble anywhere in the world... It seems to me that their offer to return to India as a group is a unique one and one that should by all means be accepted and acted upon promptly.  An offer like the present one comes only rarely in the history of scientific development of a nation which, scientifically is obviously coming of age.''

Oort, who was well known as being more reserved in evaluating new opportunities, nevertheless praised the proposal on 23 October 1961: ``Their plans are reasonable and balanced and appear extremely suitable, starting an active centre of radio-astronomical research in your county; I should therefore like to support their proposal whole-heartedly.  You are fortunate in having a group of young Indians who appear to be so well-equipped to start a research centre in a subject that is at present in the centre of interest.  There is no doubt in my mind that your country... would profit if you could succeed in realizing their plans... ''

During the opening remarks at the conference ``The Metre Wavelength Sky'' on 9 Dec 2013, the TIFR Director, Professor Mustansir Barma, discussed the founding of the TIFR radio astronomy group by Bhabha in 1961-1963. Prof.Barma discussed correspondence from late 1961-early 1962 between Bhabha and Mark Oliphant, Director of the Research School of Physical Sciences at the Australian National University - ANU.  Oliphant was a former colleague and friend from the Cavendish Laboratory in the 1930s.  Oliphant had heard about the proposed Indian radio astronomy project from his colleague Bart Bok, the Director of the Mt Stromlo Observatory of ANU.  On 6 October 1961 Oliphant wrote:
``You will be aware of the great reputation which Australia has in radioatronomy as a result of the excellent work of a relatively small group of men in the Division of Radiophysics of CSIRO.  A 210 foot radiotelescope is being opened at Parkes on 31st October [1961]... [Then a discussion of other new projects in Australia followed: the new Mills Cross and Hanbury Brown's intensity interferometer.]...  We find the presence in Australia of strong groups in radioastronomy to be most stimulating to us all and very attractive to research students.  I feel sure that Indian science would be richer if some means can be found to establish radioastronomy as a vigorous branch of research and teaching.  We in Australia would welcome the presence of colleagues in this field in neighbouring India and would be happy to help and collaborate in every possible way\footnote{TIFR archives. I am indebted to Ms. Oindrila Raychaudhuri of the TIFR Archives (Mumbai) for providing this letter.}.''

On 10 November 1961, T.K. Menon was invited to meet Bhabha for two hours in Washington, D.C. to discuss various aspects of the proposal from Krishnan et al\footnote{Report of this discussion, written 14 November 1961 by T.K.Menon, sent to Swarup.}. Menon wrote ``... he is very enthusiastic and wants to build up a large and effective group as part of the Tata Institute of Fundamental Research.'' The possible academic positions for the four scientists were discussed as well as possible budgets for the proposed radio telescopes.  For the first solar project, a budget of 5 lakhs\footnote{1 lakh = 100,000.} or about \$100,000 US in 1961 was suggested.  ``Then he talks quite happily of final outlays of 50 to 100 lakhs (\$1 to 2 million US).  Unbelievable for us at present.  But he seemed perfectly sincere and credible. .. he impressed me with his earnestness and wanting to make quick decisions... In general he had very little good [sic] to say about UGC [Universities Grants Commission] and CSIR and particularly NPL. ...As you can see I am much encouraged by the whole discussion and I hope you are too. ...Bhabha suggested the major project could be put anywhere in India suitable for us. Suggested places Bombay, Ootacamund, Kodaikanal, Poona or anywhere else we want.''

\subsection{Planning for a Radio Astronomy Group in India: 1962-1963}
A few months followed with no response from Bhabha.  A key event occurred in mid-January 1962, while Bark Bok was visiting TIFR for a two week lecture tour at the invitation of Bhabha.  Bhabha was in the process of making a positive decision to offer the four radio astronomers positions at TIFR; the offer was sent as a telegram  a few days later on 20 January to Swarup and presumably the others.  ``Reference proposal from Menon, Kundu, Krishnan and yourself forwarded by M.G.K. Menon [Dean of the TIFR Physics Faculty] in September [sent to Bhabha in Europe].  We have decided to establish radioastronomy group.  Letter follows with offer. Bhabha''\footnote{Venkataraman in {\it Bhabha and his Magnificent Obsessions} (1994) has remarked that this prompt process of starting a new Tata Institute disciple represented: ``A prompt action indeed by any standards in the world!''}. Clearly, Bhabha had discussed in some detail the radio astronomy plan with Bok.  Bok wrote two letters that were posted to Swarup and to T.K.Menon, with the request that the latter letter be sent to the other three.  The first, hand written letter to Swarup was dated 21 January:  ``I wanted to write you a brief personal note right away to tell you that I think the four of you would be very wise to accept Bhabha's offer. ...I have found a most stimulating community scientifically and socially- and I am sure that you and your wife will fit in beautifully and like it here.  I hope it will work out as planned.''  The longer letter to Menon (and the other three) expanded on this theme: ``This is one of the most important letters I shall write- or have written- for a long time, because it concerns the future of a dear friend [T.K. Menon, who had been Bok's PhD student] and a group of radio astronomers of great promise.  I am writing it only after having given full consideration to the pros and cons of the situation that you will be facing if you accept the posts at the Tata Institute- following incidentally two luncheons between just Dr.Bhabha and myself.  It is my considered opinion that the offers that [he] is mailing to you and your three colleagues are as fine offers as one could hope to receive from anywhere in the world as far as research possibilities are concerned and that in matters of salary and rank he has been as fair as can be - trying to push all of you without upsetting the balance of his institute.  My recommendation is that you should accept without qualification and that neither you nor your colleagues will regret the decision.  I have been favourably impressed with the Tata Institute, it's [sic] research outlook and opportunities, it's [sic] young staff and their leaders, especially with Dr.Bhabha and Prof. M.G. K. Menon.  If I were at your stage in a career I would not hesitate to accept an offer.  The conditions for development [at the institute] are excellent in every way and the Institute has obviously the support from the government, markedly so from the Prime Minister... . Within reason Dr.Bhabha [age 52] seems to be able to get anything he asks the government for... . My advice to you and the others is to make your headquarters here at the Institute in Bombay... then set up your field station away from disturbances ..  I do not advise your [sic] going from the start to either Kodaikanal or Ootie [sic, he meant Ooty or Ootacamund- the eventual site of the Ooty Radio Telescope completed in early 1970]\footnote{Bok also described meeting a number of young TIFR faculty and their wives, e.g. B.V. Sreekantan, whom T.K. Menon knew from Bruno Rossi's (1905-1993)  cosmic ray group at MIT, and his wife, an accomplished musician.  All were to emerge as scientific leaders in the next decades: Devendra Lal (1929-2012), director of the Physical Research Laboratory from 1972-1983, S. Naranan (cosmic ray research at the Kolar gold mine) and U.R. Rao (1932-), later Chairman Space Commission and Secretary Department of Space starting in 1985.}.

On March 2 1962 Swarup wrote Pawsey, who was on his way to the National Radio Astronomy Observatory in Green Bank West Virginia, USA, as the Director.\footnote{Pawsey was the newly appointed new Director of the National Radio Astronomy Observatory (offered to Pawsey on 31 October 1961 by Prof. I.I. Rabi [1898-1988], the President of Associated Universities, Inc).  After traveling from Australia to the US in March 1962 for a one month reconnaissance visit at NRAO in Green Bank West Virginia, he arrived at NRAO about 18 March. Within a few days (circa 23 March 1962), the paralysis of his left arm and leg due to a massive glioblastoma multiforme (cancerous tumour of the brain) occurred.  After major surgery at Massachusetts General Hospital in Boston on 16 May 1962, he and his wife (accompanied by Paul Wild) arrived back in Australia from the US on 29 July 1962. He rapidly deteriorated after a few months. He died on 30 November 1962 at age 54. The original itinerary for Pawsey's trip in 1962 had included a visit to India in April 1962, during his planned return to Australia from the US via Europe and India. Likely he planned to visit Homi Bhabha in Bombay; due to his illness, he did not meet Bhabha.}  Swarup informed Pawsey about the current status of the new radio astronomy group in India.  ``This is a very pleasing offer because of the excellent research facilities available at the Tata Institute.  The promise of financial and other supports by [Bhabha] is better than we can expect elsewhere in India and besides he has a good reputation for organizing many successful groups in India... I have no doubt that each one of the four of us can alone accomplish a fair deal in Radio Astronomy in India.  But, I am hoping that you, [plus Bok and Bhabha] will make an attempt to bring us together.  We are all deeply grateful to you for recommending our proposal.  A good part of my enthusiasm in this venture arises because of your continued interest in it.''  Due to Pawsey's severe illness (which began a few weeks later), he was to play little or no role in further deliberations; no response to Swarup from Pawsey has been found in the archives.  As we have seen, Bok was already playing a role as an outside facilitator; as Pawsey was to fade from the scene after March 1962, Bok was in frequent contact with the group of four.  On 8 February 1962, Swarup wrote a full report to Bok (similar in content to the letter to Pawsey), describing the status in a positive light, with the exception of the major concerns due to the discrepant academic ranks and salaries suggested to the four radio astronomers by TIFR.

In the course of 1962, the deliberations between TIFR (Bhabha and M.G.K. Menon) and the four radio astronomers became confused due to nature of the offer from Bhahba in January.  These complex negotiations would continue for over a year.  This offer consisted of a ``Readership'' (the academic rank) for T.K. Menon, ``Fellow'' for Swarup and Kundu and ``Research Fellow'' for T. Krishnan.  The group of four had previously agreed among themselves that maintaining a small difference in rank and salaries was important in order ``to promote efficient cooperation in the beginning''.  The difference in salaries was almost a factor of two.  Pawsey's prediction of the problem of ``lack of communications'' (letter to Swarup, 18 April 1961) had been fulfilled\footnote{In Swarup's letter to Pawsey of 2 March 1962, he wrote: ``This [the discrepant ranks and salaries] is a very tricky problem, and I think that this situation has resulted partly because of our own fault. We should have dealt with this question with Professor Bhabha more clearly in the earlier stages. I have no complaints about [T.K.] Menon's offer because I think he deserves it. But, I am disappointed to see the offer to Krishnan.''}.  Fortunately, Pawsey's prediction of ``jealousies'' regarding the leadership of the group was not maintained.  Swarup and T.K. Menon accepted the offer from Bhabha (Swarup on 8 February 1962), while Mukul Kundu was initially quite displeased with the disparity of the offers.  He did finally accept the offer in mid April 1962.

Throughout the following year many letters were exchanged between T.K. Menon, Swarup and Kundu, including a number of helpful letters from Prof. M.G. K. Menon.  By the end of the year 1962- early 1963, both Kundu and Swarup had received ``Readership'' rank for their initial appointments.  Kundu had in fact visited TIFR in Bombay on a round-the-world tour in August 1962, after visiting solar radio astronomy colleagues in Australia (Kundu was offered a position in Australia, which he declined).  Kundu had moved from the University of Michigan to Cornell (an offer from T. Gold) in 1962.  Kundu's arrival to join the radio astronomy group at TIFR was delayed by a few years.  In early 1965, he arrived at TIFR, staying for three years.   Swarup (2006) has written: ``[Kundu] contributed a great deal to the growth of the group during its critical formation years.''  Mukul Kundu (1930-2010) returned to the US in 1968 to the University of Maryland, starting an active solar group, where he worked until his death in June 2010.

In early 1962 when Bhabha made the first offers to the radio astronomers, the expectation was that T.K. Menon would be the most senior member of the radio astronomy group\footnote{For example T.K. Menon was active in planning for and even buying some radio astronomy equipment in the US during the course of 1962.}. Throughout most of the year, he , Swarup and Kundu wrote several versions of the ``Proposed Program of Tata Institute Radio Astronomy Group for the Year 1963- Tata Institute Radio Astronomy Group Memo No.1'', which was submitted to Bhabha and M.G. K. Menon.  The first draft was written by Swarup on 20 July 1962, followed by revisions by Kundu on 1 August, followed by a final re-write carried out by T.K. Menon.  The memo was sent about 4 September 1962 to Bhabha and M.G.K. Menon by T.K. Menon. The memo contained the plans for the interim solar observatory at Kalyan as a phase one project. ``Details of the next phase of our program will be decided after about six months of our joining the project.'' Earlier, Swarup had written two preliminary planning documents in 1962: ``Some Interesting Radio Astronomy Projects for India'' from May 1962 and  a detailed letter to Kundu on 27 April 1962; the latter two documents contained a brief description of a large 5 km Mills Cross working at 20-25 MHz with an angular resolution of about 10 arc min, an unprecedented resolution at a wavelength of 12m.  In addition various mm solar instruments working at 8 mm were discussed to study solar flares and the slowly varying component of solar emission.  Neither instrument was to be constructed in the mid-1960s.

In November 1962, T.K. Menon told Swarup that due to family reasons he would have to postpone his arrival to India by at least a year.  However, it would be almost a decade later that he became a member of the TIFR radio astronomy group in Bombay from 1970-1974.  T. K. Menon then moved to the University of British Columbia in Canada where he remained an active astronomer until retirement.  Swarup (2010) has written:'' ..contributions by both [Menon and Kundu] were very important to the growth of the radio astronomy  [group] of TIFR in its early years.''\footnote{In 1963, Swarup was disappointed that Menon had not joined with him at TIFR during the crucial year 1963. He wrote Bok on 30 August 1963 from Bombay: ``Kochu's [T.K. Menon] decision for not coming to India disappointed me considerably, but I am hoping that he will change his mind after seeing concrete evidence of our progress. I had a talk with him last December [1962] in the USA. I found that he was afraid of work conditions in India. My experience[s] during the last few months have proved that his [doubts] are unreasonable.'' }

Also during 1962 the TIFR negotiations with T. Krishnan broke down\footnote{Already on 1 December 1961, Swarup wrote T.K. Menon, sounding an alarm about the chances of T.Krishnan joining the group at TIFR. ``Unfortunately, we were not able to grow common views with him at the time of the IAU [August 1961].''}. Swarup had mentioned to Bok on 8 February 1962: ``I have also written a long letter to Krishnan [still in Australia at CSIRO] but I have my doubts if he will join us in view of the considerably lower offer that had been made to him.'' Krishnan wrote to Swarup on 9 March 1962: ``Thank you very much for your appreciated letter.  My reactions to Bhabha's offer are quite honestly contained in my letters to him and Kochu [Menon].  However, the whole background is such that it is almost an unacceptable offer.''  Krishnan was in the US until the early 1970s, working with Ron Bracewell at Stanford.  He then moved to Chennai in India, becoming the Director of the Madras Institute of Technology from 1971 to 1975. 

Thus in the beginning of 1963, only Swarup was available to start the radio astronomy group.  A major motivation was his loyalty to India.  On 15 November 1962 he wrote to Mukul Kundu, who was planning to postpone his arrival in India during the course of 1963: ``I am confident that we can establish one of the strongest group[sic] in India in a few years particularly if we can work and plan together with confidence and concentrated effort.  I assure you that I plan to return to India for good.'' Likely his intense loyalty to India had its beginning during his childhood at age 13 in 1942.  His experience during a childhood participation in a ``Quit India'' demonstration in his hometown of Thakurdwara (Uttar Pradesh) had a lasting impact.  During the demonstration the authorities fired weapons at the demonstrators.  ``And I remember there was firing and I got scared...I still remember the incident.  Because I was surprised that someone could fire at school children.  Perhaps this incident, and also Premchand's novels [A famous early 20th century Hindi language author] made me nationalistic.''\footnote{From TIFR Oral History Archives, interview with Govind Swarup conducted by Indira Chowdhury in the period 17 January 2005 to 10 February 2005. The 139 page interview was provided by Ms. Oindrila Raychaudhuri of TIFR Archives.}

As Swarup has pointed out (Swarup, 2006), the late V.K. Kapahi (1944-1999) and J.D. Isloor joined the group in August 1963 from the Atomic Energy Establishment Training School (AEET), with R.P. Sinha and D. S. Bagri following in August 1964.  All of these participated in the new construction and use of the Kalyan solar grating interferometer.  Also during 1964, N.V. G. Sarma and M.N. Joshi left NPL in New Delhi, becoming key members of the TIFR radio astronomy group.  S.H. Damle and T. Velusamy  joined the group in 1965.  In 1966, V. Balasubramanian from AEET and S. Ananthakrishnan from the Institute of Radio Physics and Electronics of the University of Calcutta also joined the TIFR group; the former worked on the phase shifters and the later on the electronics and control system of the Ooty radio telescope.

\section{Conclusion }

\begin{figure}[h!]
\centerline{\includegraphics[width=5cm]{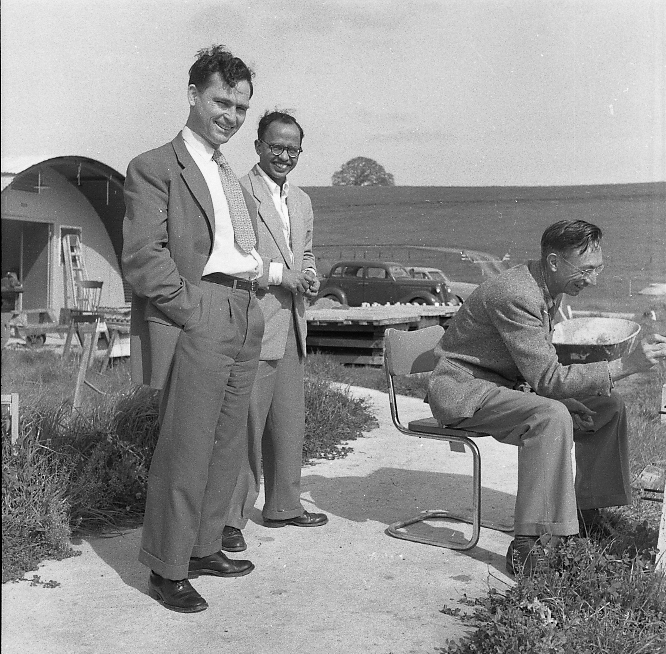}}
\caption{ Govind Swarup, R. L. Bracewell and J. L. Pawsey at the Stanford
Heliopolis site circa 1958. \label{f:five}}
\end{figure}

Swarup, Bracewell and Pawsey are shown (Fig. 5) at the Stanford Heliopolis site in circa 1958, while Pawsey was chiselling his signature on one of the telescope piers of the 11 cm spectroheliograph. 

In 1963, the stage was set for a remarkable scientific transformation in India as the TIFR radio astronomy group was at the threshold of rapid growth.  H.J. Bhaba anticipated this in a letter to Govind Swarup on 3 April 1962: ``I am very pleased at starting this project, and if your group fulfils the expectations we have of it, this could well lead to some very much bigger equipment and work in Radio Astronomy in India than we foresee at present.''  Bhabha's farsighted vision has certainly been fulfilled, 52 years later.

The ``networking'' between Indian, British, Australian and American astronomers that lead to the formation of TIFR radio astronomy (now the National Centre for Radio Astrophysics) shows that some scientific traditions of the former British Empire survived into the mid-20th century.   Fortunately for the process in 1961-62, Homi Bhabha at the TIFR was receptive to the initiative of the ex-pat Indian astronomers in the US.  Bhabha, with the support of M.G.K. Menon, took a chance on the 32 year old Govind Swarup in 1961, even though he had little firsthand experience with ionospheric physics or radio astronomy.  Pawsey, Oort, Bok and Oliphant verified the credentials of Swarup and the others.  Collaboration between many groups accelerated the advancement of radio astronomy as TIFR had its modest beginning with the solar radio telescope at Kalyan after 1965.  Two of the largest radio telescopes (the Ooty Radio Telescope and the Giant Metre Radio Telescope) ever constructed grew out of this endeavour.

Why did the process- the formation of a new radio astronomy group- work so efficiently?  Key components were, of course, having competent, experienced young scientists present at the right time in the US, Australia and Europe, all interested in returning to India.  Good connections between generations were also important, with trusting communications between them.  ``Networking'' led to contacts with senior astronomers who had been colleagues and friends of H.J. Bhabha.  Bhabha made rapid decisions; he knew how to find the ``right person'' with energy and competence.

In the next decades, the TIFR radio astronomers would create two unique radio telescopes: the Ooty Radio Telescope (1970) and later the GMRT (2000),  see Swarup (2006).The foundations for the creation of a radio astronomy institute that has remained a world leader in radio astronomy had been laid by Govind Swarup and a group of gifted individuals.  Their creations have thrived for over 50 years.  
The saga of the formation of TIFR radio astronomy is an example of learning from the past as we prepare for the future: 
``History is for human self-knowledge... .  the only clue to what man can do is what man has done.  The value of history, then, is that it teaches us what man has done and thus what man is.'' (R.G. Collingwood, 1889-1943)

\section*{Acknowledgements} The National Radio Astronomy Observatory is a facility of the National Science Foundation operated under a cooperative agreement by Associated Universities, Inc.  I thank Govind and Bina Swarup for their hospitality during many visits to India since 1970.  Our two families have shared many experiences.  Govind has provided advice over the last few years concerning the connections between Pawsey and TIFR radio astronomy.  In December 2013, he generously provided me with access to his extensive archive at NCRA.  Helpful comments have been provided by G. Swarup, A.M. Goss, S. Ananthakrishnan, Rajaram Nityananda, Harry Wendt, R.D. Ekers , R.H. Frater and Sivasankaran Srikanth.   I would like to extend  my especial thanks to Jayaram Chengalur, who made this contribution possible.
%------------------------------------------------------------------------------%

%------------------------------------------------------------------------------%
% bibliography: produced from ADS using custom format of                       %
%                                                                              %
%     %z132 \\bibitem[%\2%(y)%\3m]%{R}\n   %\8.1g,%\Y,%\q,%\V,%\ p             %
%------------------------------------------------------------------------------%

\label{lastpage}
%------------------------------------------------------------------------------%
\end{document}